\pdfoutput=1

\documentclass[aps,pra,10pt,twocolumn,showpacs,superscriptaddress,footnoteinbib]{revtex4}
\usepackage{amsmath}
\usepackage{latexsym}
\usepackage{amssymb}
\usepackage{graphics,epstopdf}
\usepackage[colorlinks=false, citecolor=blue, urlcolor=blue ]{hyperref}

\begin{document}
\title{Hardy's nonlocality argument as witness for post-quantum correlation}

\author{Subhadipa Das}
\email{sbhdpa.das@bose.res.in}
\affiliation{S.N. Bose National Center for Basic Sciences, Block JD, Sector III, Salt Lake, Kolkata-700098, India}

\author{Manik Banik}
\email{manik11ju@gmail.com}
\affiliation{Physics and Applied Mathematics Unit, Indian Statistical Institute, 203 B.T. Road, Kolkata-700108, India}

\author{Ashutosh Rai}
\email{arai@bose.res.in}
\affiliation{S.N. Bose National Center for Basic Sciences, Block JD, Sector III, Salt Lake, Kolkata-700098, India}

\author{MD Rajjak Gazi}
\email{rajjakgazimath@gmail.com}
\affiliation{Physics and Applied Mathematics Unit, Indian Statistical Institute, 203 B.T. Road, Kolkata-700108, India}

\author{Samir Kunkri}
\email{skunkri@yahoo.com}
\affiliation{Mahadevananda Mahavidyalaya, Monirampore, Barrakpore, North 24 Parganas-700120, India}

\begin{abstract}
Recently, Gallego \emph{et.al.} [\href{http://dx.doi.org/10.1103/PhysRevLett.107.210403}{Phys. Rev. Lett  {\bf107}, 210403 (2011)}] proved that any future information principle aiming at distinguishing between quantum and post-quantum correlation must be intrinsically multipartite in nature.  We establish similar result by using device independent success probability of Hardy's nonlocality argument for tripartite quantum system. We construct an example of a tri-partite Hardy correlation which is post-quantum but satisfies not only all bipartite information principle but also the GYNI inequality.
\end{abstract}


\maketitle

\section{Introduction}
Recently, understanding the correlations among distant observers which are compatible with our current description of nature based on quantum mechanics has generated much interest. It has been shown that, some general physical principles can restrict the set of no-signalling correlations among distant observers to a significant degree. Information theoretic principles like information causality \cite{pawlowski} and nontrivial communication complexity \cite{vandam1,vandam2} are novel proposals to single out the quantum correlations, from rest of the no-signalling correlations, when two distant observers (or in some cases even when more than two observers \cite{yang1}) are involved. However, by applying the known information principles to the bipartite case, it has not been possible to derive the full set of quantum correlations resulting from the Hilbert space structure of quantum mechanics.

For a multipartite (more than two subsystems) scenario, the situation becomes even more complex and extremely challenging. Very recently, interesting results \cite{gallego,yang1} have been produced when more than two distant observers are involved. In \cite{gallego}, Gallego \emph{et.al.} provide an example of tripartite no-signalling correlation which is \emph{time-ordered-bi-local} (TOBL) \cite{pironio,gallego1,barrett} and therefore respects any bipartite information principle, yet this correlation is nonquantum (unphysical) since they violate the \emph{guess-your-neighbor's-input} (GYNI) inequality \cite{almeida}. Thus, this result demonstrates that in general any bi-principle is insufficient for deriving the set of all multi-party physical (quantum) correlations. A similar example is also provided in the work by Yang \emph{et.al.}\cite{yang1}, where an extremal point of the tripartite non-signaling polytope is proved to be non-quantum. The non-quantumness of this correlation is shown through a violation of an inequality (eqn.-(A1) in \cite{yang1}) which is satisfied by all quantum correlations; however, in contrast with \cite{gallego}, this correlation respects the GYNI inequality.

In the present work, we give a TOBL correlation which is non-quantum since it exceeds the maximum success probability of Hardy argument for tripartite quantum correlations. To show this, first we prove that in quantum mechanics the maximum success probability of Hardy argument for a tripartite system is $0.125$, which was earlier known only for projective measurement on three qubits system \cite{ghosh}. Then we explicitly construct a tripartite correlation in a general probabilistic theory with the following properties: (i) the correlation is TOBL and hence satisfies all bi-partite information principle, (ii) the correlation shows Hardy nonlocality with success probability taking the value $0.2$ which is greater than the maximum value $0.125$ that can be achieved within quantum mechanics.
Interestingly, this correlation respects all the GYNI inequalities, indicating that any multipartite information principle based on GYNI type of games may still be insufficient in distinguishing physical correlations.

The paper is organized as follows. In section (\ref{hardy}) we prove that the device independent success probability for Hardy nonlocality for tripartite quantum system is $\frac{1}{8}$. In section (\ref{tobl}), we briefly described the TOBL correlations. In section (\ref{nonquantum}), we present a tripartite no signaling probability distribution which belong to TOBL set with success probability for Hardy nonlocality argument greater than that for quantum system, and finally in section (\ref{conclusion}) we give our conclusions.
\section{Tripartite Quantum Systems and Hardy Argument }\label{hardy}
Lucian Hardy first provided an argument which reveals nonlocality within quantum mechanics without using any inequality \cite{hardy}. In order to present Hardy's argument in the general probabilistic theories, we consider the set of tripartite no-signalling correlations with binary input and binary output for each party---the set of such correlation are known to be points of a polytope in a $26$-dimensional space with $53,856$ extremal points \cite{pironio}. A tripartite two-input-two-output Hardy correlation is defined by some restrictions on a certain choice of $5$ out of $64$ joint probabilities in the correlation matrix \cite{kar,cereceda,boschi,kunkri}. The following five conditions, for example, define a tripartite Hardy correlation:

\begin{eqnarray}
 P(A_1=+1, ~B_1=+1, ~C_1=+1)> 0 \nonumber \\
 P(A_2=+1, ~B_1=+1, ~C_1=+1)=0 \nonumber\\
 P(A_1=+1, ~B_2=+1, ~C_1=+1)=0 \\
 P(A_1=+1, ~B_1=+1, ~C_2=+1)=0 \nonumber\\
 P(A_2=-1, ~B_2=-1, ~C_2=-1)=0 \nonumber
\end{eqnarray}
where, $\{A_1,A_2\}$, $\{B_1,B_2\}$ and $\{C_1,C_2\}$ are outcome of respective local observables corresponding to measurements performed by three distant parties, say Alice, Bob and Charlie, and $\pm1$ are the possible measurement outcomes. One can easily show that the above correlation can not be reproduced by any local realistic model.

The joint probability appearing in the first condition, inequality (1), is the success probability for Hardy's nonlocality argument. In quantum mechanics, for three qubits systems subjected to local projective measurements, the maximum value of the success probability of Hardy argument has been shown to be $0.125$ \cite{ghosh}. In view of a recent result providing a device independent bound for the success probability of a bipartite Hardy argument \cite{scarani}, one can ask what is the maximum probability of success for Hardy nonlocality for tripartite quantum systems. We extend this result for the tripartite case \cite{proof}:

 \textbf{Proposition}: \emph{The maximum value of the success probability of the Hardy argument for tripartite quantum systems is $\frac{1}{8}$.}

\textbf{Proof:} To prove the proposition we use an important lemma in \cite{masanes}, which states that, given four projectors  $P$, $I-P$, $Q$, $I-Q$ acting on a Hilbert space $\mathbb{H}$, there exists an orthogonal basis which allows to decompose $\mathbb{H}$ as a direct sum of subspaces $\mathbb{H}^i$ of dimension $d\leq2$ for each $i$, such that each of the four projectors can be written as $\Pi=\bigoplus_i \Pi^i$, where $\Pi^i$ acts on $\mathbb{H}^i$ and $\Pi\in\{P, I-P,Q, I-Q\}$.

In quantum mechanics joint probabilities for the outcomes of measurements on a tripartite system are given by $$P(A,B,C|\hat{a},\hat{b},\hat{c})=Tr(\rho M_{A|\hat{a}}\otimes M_{B|\hat{b}}\otimes M_{C|\hat{c}})$$ where $\rho$ is the state of the system and $M_{A|\hat{a}}$, $M_{B|\hat{b}}$, $M_{C|\hat{c}}$ are the measurement operators associated to outcomes $A$, $B$, $C$ of measurements $\hat{a}$, $\hat{b}$, $\hat{c}$ respectively. These measurements operators are POVM, in general; but as we are not restricting the dimension of the Hilbert space, without loss of generality Neumark's theorem allows us to consider only projective measurements. Each of Alice, Bob and Charlie choose to perform from two measurements, say $\hat{a}=\{\hat{a}_1,\hat{a}_2\}$, $\hat{b}=\{\hat{b}_1,\hat{b}_2\}$ and $\hat{c}=\{\hat{c}_1,\hat{c}_2\}$ respectively with measurement results $\pm1$ for each measurements. Now, $\hat{a}_1=M_{+1|\hat{a}_1}-M_{-1|\hat{a}_1}$ and $\hat{a}_2=M_{+1|\hat{a}_2}-M_{-1|\hat{a}_2}$, where $M_{\pm1|\hat{a}_{1(2)}}$ are projection operators. According to the above mentioned lemma \cite{masanes}, it follows that $M_{A|\hat{a}}=\bigoplus_iM^i_{A|\hat{a}}$ for all $A$ and $\hat{a}$. Let us denote $M^i=M^i_{+1|\hat{a}}+M^i_{-1|\hat{a}}$ for all $\hat{a}$. Also this lemma is valid for Bob's and Charlie's end. Using analogous notations for Bob's and Charlie's operators, we can write
\begin{eqnarray}
P(A,B,C|\hat{a},\hat{b},\hat{c})&=&\sum_{i,j,k}q_{ijk}Tr(\rho_{ijk}M^i_{A|\hat{a}}\otimes M^j_{B|\hat{b}}\otimes M^k_{C|\hat{c}})\nonumber \\
&=&\sum_{i,j,k}q_{ijk}P_{ijk}(A,B,C|\hat{a},\hat{b},\hat{c})
\end{eqnarray}
where, $q_{ijk}=Tr(\rho M^i_{A|\hat{a}}\otimes M^j_{B|\hat{b}}\otimes M^k_{C|\hat{c}})$ and $\rho_{ijk}=(M^i\otimes M^j \otimes M^k \rho M^i\otimes M^j \otimes M^k)/q_{ijk}$ is a three qubit state. As $q_{ijk}\geq0$ for all $i,j,k$ and $\sum_{i,j,k}q_{ijk}=1$ the Hardy constraints (last four condition of eqn. (1)) are satisfied for $P$ if and only if they are satisfied for each of the $P_{ijk}$. Which implies that
\begin{eqnarray}
P(A_1=+1, ~B_1=+1, ~C_1=+1)\nonumber\\=\sum_{i,j,k}q_{ijk}P_{ijk}(A^{i}_1=+1, ~B^{j}_1=+1, ~C^{k}_1=+1)
\end{eqnarray}
Being a convex sum, the success probability Hardy argument is therefore less or equal to the largest element in the combination, which is $\frac{1}{8}$.~~~~~~~~~~~~~~~~~~~~~~~~~~~~~~~~~~~$\Box$

In the remaining part of this paper, for convenience we adopt another notation for the joint correlation probabilities expressed as  $P(abc|xyz)$ with input $x,y,z\in\{0,1\}$ and output $a,b,c\in\{0,1\}$.

\section{Time-ordered Bi-local(TOBL) Correalions}\label{tobl}

A tripartite no-signaling probability distribution  $P(abc|xyz)$ belongs to TOBL \cite{pironio,gallego1,barrett} if it can be written as
\begin{eqnarray}
P(abc|xyz)=\sum_{\lambda}p_{\lambda}P(a|x,\lambda)P_{B \rightarrow C}(bc|yz,\lambda)\\
          =\sum_{\lambda}p_{\lambda}P(a|x,\lambda)P_{B \leftarrow C}(bc|yz,\lambda)
\end{eqnarray}
and analogously for $B|AC$ and $C|AB$, where $p_{\lambda}$ is the distribution of some random variable $\lambda$, shared by the parties. The distributions $P_{B \rightarrow C}$ and $P_{B \leftarrow C}$ respect the conditions
\begin{eqnarray}
P_{B \rightarrow C}(b|y,\lambda)=\sum_{c}P_{B \rightarrow C}(bc|yz,\lambda)\\
P_{B \leftarrow C}(c|z,\lambda)=\sum_{b}P_{B \leftarrow C}(bc|yz,\lambda)
\end{eqnarray}
From these equations it is clear that the distributions $P_{B \rightarrow C}$ allow signaling from Alice to Bob and $P_{B \leftarrow C}$ allow signaling from Bob to Alice.
If a tripartite no-signaling probability distribution $P(abc|xyz)$ belongs to the set of TOBL distributions, all possible bipartite distributions derived by applying any local wirings on $P(abc|xyz)$ are local, i.e., the probability distribution $P(abc|xyz)$ respects all bi-partite physical principle \cite{gallego1}.

\section{non quantum correlation satisfying bipartite principle}\label{nonquantum}
Now we show that, there exist a tripartite Hardy correlation which lies in the TOBL set. Since, this correlation belongs to the TOBL set, it must satisfy all bipartite information principle. The probability distribution is given in the TABLE-I.
\begin{table}[t]
\centering
\caption{Tripartite no-signaling probability distribution $ P(abc|xyz)$ with Hardy's success $1/5$.}
\begin{tabular}{lllllllll}
\hline \hline
$ xyz\backslash abc$  & ~~$000$ & ~~$001$ & ~~$010$ & ~~$011$ & ~~$100$ & ~~$101$ & ~~$110$ & ~~$111$ \\
\hline \\
~~$000$ & ~~~$\frac{1}{5}$ & ~~~$0$ & ~~~$0$ & ~~~$\frac{1}{5}$ & ~~~$0$ & ~~~$\frac{1}{5}$ & ~~~$\frac{1}{5}$ & ~~~$\frac{1}{5}$\\\\
~~$001$ & ~~~$0$ & ~~~$\frac{1}{5}$ & ~~$\frac{1}{10}$ & ~~$\frac{1}{10}$ & ~~$\frac{1}{10}$ & ~~$\frac{1}{10}$ & ~~~$\frac{2}{5}$ & ~~~$0$\\\\
~~$010$ & ~~~$0$ & ~~$\frac{1}{10}$ & ~~~$\frac{1}{5}$ & ~~$\frac{1}{10}$ & ~~$\frac{1}{10}$ & ~~~$\frac{2}{5}$ & ~~$\frac{1}{10}$ & ~~~$0$\\\\
~~$011$ & ~~~$0$ & ~~$\frac{1}{10}$ & ~~$\frac{1}{10}$ & ~~~$\frac{1}{5}$ & ~~~$\frac{2}{5}$ & ~~$\frac{1}{10}$ & ~~$\frac{1}{10}$ & ~~~$0$\\\\
~~$100$ & ~~~$0$ & ~~$\frac{1}{10}$ & ~~$\frac{1}{10}$ & ~~~$\frac{2}{5}$ & ~~~$\frac{1}{5}$ & ~~$\frac{1}{10}$ & ~~$\frac{1}{10}$ & ~~~$0$\\\\
~~$101$ & ~~~$0$ & ~~$\frac{1}{10}$ & ~~~$\frac{2}{5}$ & ~~$\frac{1}{10}$ & ~~$\frac{1}{10}$ & ~~~$\frac{1}{5}$ & ~~$\frac{1}{10}$ & ~~~$0$\\\\
~~$110$ & ~~~$0$ & ~~~$\frac{2}{5}$ & ~~$\frac{1}{10}$ & ~~$\frac{1}{10}$ & ~~$\frac{1}{10}$ & ~~$\frac{1}{10}$ & ~~~$\frac{1}{5}$ & ~~~$0$\\\\
~~$111$ & ~~~$\frac{2}{5}$ & ~~~$0$ & ~~~$0$ & ~~~$\frac{1}{5}$ & ~~~$0$ &  ~~~$\frac{1}{5}$ & ~~~$\frac{1}{5}$ & ~~~$0$ \\\\
\hline \hline
\end{tabular}

\label{TABLE I}
\end{table}
In this table, $ P(000|000)=\frac{1}{5},~P(000|001)=0,~P(000|100)=0,~P(000|010)=0,~P(111|111)=0$. The success probability of Hardy argument for this correlation is $\frac{1}{5}$ which is strictly larger than the maximum achievable quantum value $\frac{1}{8}$ and thus this is a nonquantum correlation.

To prove that the distribution $P(abc|xyz)$ belongs to the TOBL set we show that it admits TOBL decomposition for all bi-partition. Note that the correlation considered in the TABLE-I is symmetric under any permutation of the parties, so here it it is sufficient to provide a TOBL model in any one bipartition, say $A|BC$.

Probability distribution appearing in the TOBL decomposition for the bi-partition $A|BC$, are such that for a given $\lambda$ Alice's outcome $a$ depends only her measurement settings $x$. Also, for given $\lambda$, $P_{B \rightarrow C}(b|y,\lambda)$ is independent of $z$ but for $B \rightarrow C$, $c$ depends on both $ y$ and $z$. Similarly, for given $\lambda$, $P_{B\leftarrow C}(c|z,\lambda)$ is independent of $y$ but for $B\leftarrow C$, $b$ depends on both $ y$ and $z$. Let $a_x$, $b_y$ and $c_{z}$ denote the outcomes for Alice, Bob and Charlie for the respective inputs $x$, $y$ and $z$. In the following TABLE-II and TABLE-III , the outputs are deterministic and the weights $p_\lambda$ are same. For any given $\lambda$, the outcome assignments of $A$ in both the tables are same.
\begin{table}[h]
\centering
\caption{ TOBL decomposition for the case $A|B \rightarrow C$.}
\begin{tabular}{lllllllllll}
\hline
$ $ & $\lambda$ & ~$p_\lambda $ & ~$a_0$ & $a_1$ & $b_0$ & $b_1$ & $c_{00}$ & $c_{01}$ & $c_{10}$ & $c_{11}$ \\
\noalign{\smallskip}\hline\noalign{\smallskip}
$ $ & $1$ & ~$\frac{1}{10}$ & ~$0$ & $0$ & $1$ & $1$ & $1$ & $0$ & $1$ & $1$ \\
$ $ & $2$ & ~$\frac{1}{10}$ & ~$0$ & $0$ & $1$ & $1$ & $1$ & $1$ & $0$ & $1$ \\
$ $ & $3$ & ~$\frac{1}{10}$ & ~$1$ & $0$ & $0$ & $0$ & $1$ & $1$ & $1$ & $0$ \\
$ $ & $4$ & ~$\frac{1}{10}$ & ~$1$ & $0$ & $1$ & $0$ & $0$ & $0$ & $1$ & $0$ \\
$ $ & $5$ & ~~$\frac{1}{5}$  & ~$1$ & $0$ & $1$ & $0$ & $1$ & $0$ & $1$ & $0$ \\
$ $ & $6$ & ~$\frac{1}{10}$ & ~$0$ & $1$ & $0$ & $0$ & $0$ & $1$ & $1$ & $1$ \\
$ $ & $7$ & ~$\frac{1}{10}$ & ~$0$ & $1$ & $0$ & $1$ & $0$ & $1$ & $0$ & $0$ \\
$ $ & $8$ & ~$\frac{1}{10}$ & ~$1$ & $1$ & $1$ & $0$ & $0$ & $0$ & $0$ & $1$ \\
$ $ & $9$ & ~$\frac{1}{10}$ & ~$1$ & $1$ & $0$ & $1$ & $1$ & $0$ & $0$ & $0$ \\
\hline
\end{tabular}

\label{TABLE II}
\end{table}

\begin{table}[h]
\centering
\caption{TOBL decomposition for the case $A|B \leftarrow C$.}
\begin{tabular}{lllllllllll}
\hline
$ $ & $\lambda$ & ~$p_\lambda $ & ~$a_0$ & $a_1$ & $b_{00}$ & $b_{01}$ & $b_{10}$ & $b_{11}$ & $c_{0}$ & $c_{1}$ \\
\noalign{\smallskip}\hline\noalign{\smallskip}
$ $ & $1$ & ~$\frac{1}{10}$ & ~$0$ & $0$ & $1$ & $0$ & $1$ & $1$ & $1$ & $1$ \\
$ $ & $2$ & ~$\frac{1}{10}$ & ~$0$ & $0$ & $1$ & $1$ & $0$ & $1$ & $1$ & $1$ \\
$ $ & $3$ & ~$\frac{1}{10}$ & ~$1$ & $0$ & $1$ & $1$ & $1$ & $0$ & $0$ & $0$ \\
$ $ & $4$ & ~$\frac{1}{10}$ & ~$1$ & $0$ & $0$ & $1$ & $0$ & $0$ & $1$ & $0$ \\
$ $ & $5$ & ~~$\frac{1}{5}$  & ~$1$ & $0$ & $1$ & $1$ & $0$ & $0$ & $1$ & $0$ \\
$ $ & $6$ & ~$\frac{1}{10}$ & ~$0$ & $1$ & $0$ & $1$ & $1$ & $1$ & $0$ & $0$ \\
$ $ & $7$ & ~$\frac{1}{10}$ & ~$0$ & $1$ & $0$ & $0$ & $1$ & $0$ & $0$ & $1$ \\
$ $ & $8$ & ~$\frac{1}{10}$ & ~$1$ & $1$ & $0$ & $0$ & $0$ & $1$ & $1$ & $0$ \\
$ $ & $9$ & ~$\frac{1}{10}$ & ~$1$ & $1$ & $1$ & $0$ & $0$ & $0$ & $0$ & $1$ \\
\hline
\end{tabular}

\label{TABLE III}
\end{table}
It is interesting to observe that the correlation in TABLE-I satisfies the most general GYNI inequality \cite{almeida} (see also \cite{almeida1}) whose violation certifies non-quantumness of a correlation. Thus the post-quantum nature (which is guaranteed by the larger success probability for Hardy's argument compared to quantum result) of the correlation given in TABLE I can not be witnessed by violating the GYNI inequality as happened for the correlation given in \cite{gallego}. Therefore the nonlocality of this post-quantum correlation is qualitatively different from the nonlocality of the post-quantum correlation that appeared in \cite{gallego}.

\section{Conclusions}\label{conclusion}
Distinguishing physically realized correlations from unphysical ones by some fundamental principle is an active area of research in the foundational perspective. Rather it has been proved that nonlocality is useful resource for device independent cryptography \cite{acin}. So it is very important to know which nonlocal correlations can be obtained by physical means. Like celebrated Bell inequality \cite{bell}, the elegant argument of Hardy \cite{hardy} reveals the nonlocality of quantum mechanics. Again like the device independent value of Bell violation for bipartite quantum mechanical system i.e. Cirel'son bound \cite{cirel}, the optimal success probability for Hardy's nonlocality argument \cite{scarani} in quantum mechanics could be a potential witness for detecting post-quantum no signaling correlations. We derive the device independent success probability of Hardy's argument for tripartite quantum system. Then we provide an explicit tripartite correlation, which satisfies any bipartite information principles, but shows Hardy nonlocality with probability which is post-quantum. In this way we establish that this device independent value is a potential witness for tripartite post-quantum correlations. Our example also satisfies most general GYNI inequality indicating insufficiency of multipartite information principle based on GYNI game in identifying physical correlations. It shows the intricate structure of physically allowed correlations when more than two spatially separated observers are involved.
\begin{acknowledgments}
It is a pleasure to thank Guruprasad Kar for motivating us in this direction. We also thank Sibasish Ghosh and Ramij Rahaman for many stimulating discussion regarding Hardy's nonlocality argument. SD and AR
acknowledge support from the DST project SR/S2/PU-16/2007.
\end{acknowledgments}



\begin{thebibliography}{99}
\bibitem{pawlowski} M. Pawlowski,T. Paterek, D. Kaszlikowski, V. Scsrani, A.Winter and M. Zukowski, \href{http://dx.doi.org/10.1038/nature08400}{Nature {\bf461}, 1101 (2009).}
\bibitem{vandam1}W. van Dam, \href{http://cs.ucsb.edu/~vandam/oxford_thesis.pdf}{Nonlocality and Communication complexity, Ph.D. thesis, University of Oxford (2000).}
\bibitem{vandam2}W. van Dam, \href{http://arxiv.org/pdf/quant-ph/0501159.pdf}{arXiv:0501159 [quant-ph] (2005).}
\bibitem{gallego} R. Gallego, L. Erik Wurflinger, A. Acin and M. Navascues, \href{http://dx.doi.org/10.1103/PhysRevLett.107.210403}{Phys. Rev. Lett  {\bf107}, 210403 (2011).}
\bibitem{yang1}T. H. Yang, D. Cavalcanti, M. L. Almeida, C. Teo and V. Scarani, \href{http://dx.doi.org/10.1088/1367-2630/14/1/013061}{New J. Phys. {\bf14}, 013061 .}
\bibitem{pironio}S. Pironio, J. D. Bancal and V. Scarani, \href{http://dx.doi.org/doi:10.1088/1751-8113/44/6/065303 }{J. Phys. A:Math. Theor. {\bf44} 065303 (2011).}
\bibitem{gallego1} R. Gallego, L. Erik Wurflinger, A. Acin and M. Navascues, \href{http:10.1103/PhysRevLett.109.070401}{Phys. Rev. Lett. 109, 070401 (2012)}
\bibitem{barrett}J. Barrett, S. Pironio, J. Bancal and N. Gisin, \href{http://arxiv.org/abs/1112.2626}{arXiv:1112.2626 [quant-ph] (2011).}
\bibitem{almeida}M. L. Almeida \emph{et.al.}, \href{http://dx.doi.org/10.1103/PhysRevLett.104.230404}{Phys. Rev. Lett.{\bf104}, 230404 (2010).}
\bibitem{ghosh}S. K. Choudhary \emph{et.al.}, \href{http://arxiv.org/abs/0807.4414}{Quantum Information and Computation, {\bf Vol. 10}, No. 9 and 10 (2010) 0859–0871.}
\bibitem{hardy} L. Hardy, \href{http://dx.doi.org/10.1103/PhysRevLett.68.2981}{Phys. Rev. Lett. {\bf68}, 2981 (1992)}; L. Hardy, \href{http://dx.doi.org/10.1103/PhysRevLett.71.1665}{Phys. Rev. Lett. {\bf71}, 1665 (1993).}
\bibitem{kar}G. Kar, \href{http://dx.doi.org/10.1103/PhysRevA.56.1023}{Phys. Rev. A, {\bf56}, 1023 (1997)};
\bibitem{cereceda}J. Cereceda,\href{http://dx.doi.org/10.1016/j.physleta.2004.06.004}{ Phys. Lett. A {\bf327}, 433 (2004)};
\bibitem{boschi}D. Boschi, S. Branca, F. De Martini, L. Hardy, \href{http://dx.doi.org/10.1103/PhysRevLett.79.2755}{Phys.Rev. Lett. {\bf79}, 2755 (1997)};
\bibitem{kunkri}S. Kunkri, S. K. Choudhary, \href{http://dx.doi.org/10.1103/PhysRevA.72.022348}{Phys. Rev. A {\bf72}, 022348(2005).}
\bibitem{scarani}R. Rabelo, L. Y. Zhi and V. Scarani, \href{http://arxiv.org/abs/1205.3280v2}{arXiv:1205.3280v2 [quant-ph] (2011).}
\bibitem{proof} In fact, this result can be stated in a more general way---If an N-qubits system subjected to local projective measurements exhibit Hardy's nonlocality with the maximum success probability, say $\alpha$, then for any N-partite quantum system subjected to general dichotomic local measurements, the maximum success probability of Hardy's nonlocality argument cannot exceeds the value $\alpha$. For $N=2$ and $N=3$ the respective values of $\alpha$ are $0.09$ and $\frac{1}{8}$.
\bibitem{masanes}L. Masanes, \href{http://dx.doi.org/10.1103/PhysRevLett.97.050503}{Phys. Rev. Lett. {\bf97}, 050503 (2006).}
\bibitem{almeida1}A. Acin, M. L. Almeida, R. Augusiak and N. Brunner, \href{http://arxiv.org/abs/1205.3076v1}{arXiv:1205.3076v1 [quant-ph] (2012).}
\bibitem{acin}A. Acin, N. Brunner, N. Gisin, S. Massar, S. Pironio and V. Scarani, \href{http://dx.doi.org/10.1103/PhysRevLett.98.230501}{Phys. Rev. Lett. {\bf98}, 230501 (2007).}
\bibitem{bell}J. S. Bell, Speakable and Unspeakable in Quantum Mechanics: Collected papers on quantum philosophy, Cambridge University Press.
\bibitem{cirel}B. S. Cirel'son, \href{http://www.springerlink.com/content/l57053g573430450/}{Lett. Math. Phys. {\bf4}, 93 (1980).}
\end{thebibliography}
\end{document}